\documentstyle[prl,aps,multicol]{revtex}
\begin{document}
\tightenlines
\input{psfig}
\newcommand{\be}{\begin{equation}}
\newcommand{\ee}{\end{equation}}
\newcommand{\bq}{\begin{eqnarray}}
\newcommand{\eq}{\end{eqnarray}}
\newcommand{\bibit}{\nineit}
\newcommand{\bibbf}{\ninebf}
\def\id{{\rm 1\kern-.21em 1}}
\font\nineit=cmti9
\font\ninebf=cmbx9
\title{Spectral Properties of Coupled Bose-Einstein Condensates}
\author{Roberto Franzosi and Vittorio Penna}
\address{Dipartimento di Fisica \& Unit\`a INFM, Politecnico di Torino,
Corso Duca degli Abruzzi 24, I-10129 Torino, Italy.}
\date{\today}
\maketitle
\begin{abstract}
We investigate the energy spectrum structure of a system
of two (identical) interacting bosonic wells occupied by $N$
bosons within the Schwinger realization of the angular momentum.
This picture enables us to recognize the symmetry properties of
the system Hamiltonian $H$ and to use them for characterizing
the energy eigenstates. Also, it allows for the derivation of the
single-boson picture which is shown to be the background picture
naturally involved by the secular equation for $H$.
After deriving the corresponding eigenvalue equation, we recast it 
in a recursive $N$-dependent form which suggests a way to generate
the level doublets (characterizing the $H$ spectrum) via suitable
inner parameters. Finally, we show how the presence of doublets in the
spectrum
allows to recover, in the classical limit, the symmetry breaking effect
that characterizes the system classically.
%
\end{abstract}
\begin{multicols}{2}
\section{introduction}
After the recent observation~\cite{aemwcdaba} of
Bose-Einstein condensation in dilute atomic gases realized
by confining a macroscopic population of atoms in a potential trap
\cite{BEC}, a large amount of work has been devoted to construct
devices where two condensates~\cite{NNN} are trapped in a double-well
potential. The interaction of such coupled Bose-Einstein condensates
(BEC) gives rise to quantum phenomena such as coherent tunneling and
interference effects that have been the subject of a huge number
of studies, both theoretic~\cite{OHST} and experimental~\cite{atmdkk}.

The present paper is focussed on the dynamical aspects inherent
in the interaction of two condensates trapped in two identical wells.
Such a dynamics has been studied
thoroughly in Refs. \cite{RAGH} within the minimal interaction
scheme
\be
\cases{
&${ i\hbar {\dot \psi}_1 =
\bigl [ -\frac{\hbar^2}{2m} \triangle -v +U |\psi_1|^2 
\bigr]\psi_1 -T \psi_2 }$ \cr
& ${\-}$ \cr 
&${
i\hbar {\dot \psi}_2 =
\bigl [ -\frac{\hbar^2}{2m} \triangle -v +U |\psi_2|^2 
\bigr]\psi_2 -T\psi_1 }$ \cr}
\label{CGP}
\ee
where the classical fields $\psi_j ({\bf r}, t)$
obey two coupled Gross-Pitaevskii equations (GPE),
and $U$, $v$, $T$, describe
the interatomic scattering, the external potential and
the tunneling amplitude, respectively.
Fields $\psi_j ({\bf r})$ (often called the wave functions of
the condensate~\cite{BEC}) are defined as the expectation value
$\psi_j ({\bf r}, t)= \langle {\hat \Psi}_j ({\bf r},t) \rangle$
of the field operators ${\hat \Psi}_j$ within the many-body
quantum theory of BEC's \cite{COM1}.
The negligible space dependence of $\psi_j$ through the condensates
allows one to set $\triangle \psi_j \simeq 0$ so that
Eqs. (\ref{CGP}) reduce
to a hamiltonian system with two complex degrees of freedom
where the nonlinear cubic terms
provide the system with an ample variety
of interwell processes.

Before discussing the goals of this paper, it is useful
to briefly review the relevant traits of the dynamics issued
from the space-independent form of Eqs. (\ref{CGP}). Due to
the assumption that the bosonic wells are identical,
the associated model Hamiltonian
${\cal H}(\psi_1, \psi_2) = U(|\psi_1|^4 + |\psi_2|^4)
- v {\cal N} - T(\psi_1 \psi_2^* + \psi_2\psi_1^*)$
(${\cal N} =|\psi_1|^2 + |\psi_2|^2$)
exhibits a permutational symmetry (PS) realized by the
exchange of the dynamical variables $\psi_1$, $\psi_2$.
Physically, this is represented by the population exchange,
$|\psi_j|^2$ being the boson number of the $j$-th well
up to a volume factor.

The structure of the model phase space $\cal P$
(this is two-dimensional because $\cal N$ is a constant of
motion) reflects in a nontrivial way the presence of the PS.
For energies $E < E_*$, ($E_*$ is a critical value of the
energy depending on the model parameters) the orbits
are placed concentrically around the energy minimum, and
cover a region $C_0 \subset \cal P$ that has essentially the
structure of the harmonic oscillator phase space.
This can be shown to entail the oscillations of the two
condensates' populations around the common value ${\cal N} /2$.
Also, for $E < E_*$ the PS maps each orbit in itself.
The remaining part of $\cal P$, filled by orbits with $E> E_*$,
is formed by two (spatially) disjoint components $C_+$, $C_-$
such that ${\cal P} \equiv$  $C_+ \cup C_- \cup \,C_0$.
In this case any given energy value $E$ is associated with
two distinct orbits $\gamma_+ \in C_+$, $\gamma_-\in C_-$
such that $\gamma_{\pm} \to \gamma_{\mp}$ under the PS action.
The presence of two energy maxima (located in $C_+$ and $C_-$
symmetrically) causes such a structure.
The remarkable feature is that, when $E$ crosses $E_*$ from below,
the system undergoes a symmetry breaking (SB) phenomenon~\cite{AFKO}
(governed by a bifurcation mechanism~\cite{COM2})
since the system `must' choose to evolve
either along $\gamma_+$ or along $\gamma_-$.  
Dynamically, this entails the emergence of a permanent gap between
the condensates' populations. Such an effect is also called
a {\em self-trapping effect}~\cite{RAGH,MCWW} in that, within
a finite range of $E$, the system never leaves the region
$C_+$ ($C_-$) where was initially placed to go in $C_-$ ($C_+$).

The investigation at the quantum level of the scenario
just described has been performed in Ref. \cite{MCWW}
through the model Hamiltonian 
\bq
H = U(n^2_1 + n^2_2)
- v N - T(a_1 a_2^+ + a_2a_1^+) \, ,
\label{HQ2}
\eq
which represents the quantum counterpart of the Hamiltonian
${\cal H}(\psi_1, \psi_2)$ for the Laplacian-free
Eqs. (\ref{CGP}). A simple way to obtain $H$ relies on
the fact that for low numbers of bosons per well
(a realizable experimental situation) one can
replace the condensate wave functions $\psi_j$'s
with the raising (lowering) operator
$a_{i}$, ($a^{+}_{i}$)
obeying the canonical commutators
$[a_{i},a^{+}_{j}]=\delta_{ij}$, $i = 1, 2$.
Since $[N,H]=0$, the total number of bosons $N := n_1 + n_2$,
$n_{i} := a^{+}_{i} a_{i}$, is a constant of motion (to simplify,
we shall denote its eigenvalue by $N$ as well).  
The rigorous derivation of $H$ is effected in Ref. \cite{MCWW}
by using a mode dependent form of the field operator ${\hat \Psi}_j$
within the many-body quantum theory~\cite{COM3} of BEC's.
It is important to recall as well that model
(\ref{HQ2}) has been studied also in Ref. \cite{AFKO}
from the viewpoint of dynamical system theory.

In this paper we investigate the spectral properties
of model (\ref{HQ2}) by combining the use of the PS
and of a further symmetry involving the change $T \to -T$.
The latter will be called {\it odd symmetry} (OS)
for recalling its basic role in determining the structure of
the energy spectrum when the total boson number $N$ is odd.
Such symmetries are used extensively to show that: 

\noindent
({\it i}) after recovering the known nondegeneracy
of the Hamiltonian spectrum (see Ref. \cite{AFKO,MCWW}),
each energy eigenstate is either
symmetric or antisymmetric under the PS action,

\noindent
({\it ii}) the doublets (pairs of close energy levels
occurring in the energy spectrum when the model
parameters range in a suitable interval)
always pair a symmetric eigenstates with an
antisymmetric one; we shall show how this feature plays
a basic role in the classic limit,

\noindent
({\it iii}) the separation mechanism causing the
splitting of the energy levels (the splitting effect has been
observed numerically in Refs. \cite{AFKO,MCWW})
can be explained in a purely analytic way, 

\noindent
({\it iv}) the fact that total boson number is even/odd
dramatically
influences the eigenvalues' parity under the OS.
 
As to point ({\it i}), we wish to emphasize that the
main consequence of the nondegeneracy is to
prevent the SB phenomenon occurring in $\cal P$
as well as the ensuing self-trapping of
the system on a specific orbit of the two ones associated with
a given energy $E > E_*$.
This reflects the intrinsic tunneling effect due to the quantum
nature of the system.

For attaining results ({\it i})-({\it iv}) we first reformulate
the initial quantum problem of two coupled
(identical) wells through the Schwinger realization~\cite{ZFG}
of the spin algebra in terms of
two-boson operators. This allows one to reconstruct
Hamiltonian (\ref{HQ2}) within its dynamical algebra
(this is introduced in Appendix A).
The form of $H$ thus obtained can be interpreted in terms of
a bosonic model defined on a nonhomogeneous linear lattice with
one effective boson ({\em single-boson picture}). Such a picture 
has been derived in Ref. \cite{FPZ} and is reviewed in Sec. II
together with the related formal background.
Thanks to its group-theoretic character
such a formulation is applicable to many-well systems
with any boson number $N$. 

In Sec. III the
diagonalization of $H$ is faced in a systematic way by
making explicit the PS and OS action on the components
of the energy eigenstates. The resulting characterization
of the eigenstates leads to identify the recursive expression
of the eigenvalue equation of $H$ both for even $N$ and for odd $N$.
This, in turn, enables us to recognize explicitly
a {\it hidden} parameter able to control in an analytic
way the level distance of the doublets. Such a parameter seems
to suggest an alternative procedure to evaluate perturbatively
the energy levels.
A similar mathematical problem was analysed in Ref. \cite{AFKO} 
as to the problem of the {\em dynamical tunneling through a
separatrix}, where the splittings of the doublets were traced
by using the standard perturbation method.   

\section{spin picture of the two-well model}
A convenient way to study the spectrum of $H$ consists in
reformulating $H$ by means of the Schwinger picture of spin
operators. The latter is a two-boson realization of the
spin operators~\cite{ZFG}
$$
J_1= \frac{a_1 a_2^+ + a_2a_1^+}{2}, \,
J_2= \frac{a_1 a_2^+ - a_2 a_1^+}{2i}, \, 
J_3= \frac{n_2 - n_1}{2},$$ 
satisfying the commutators
$[J_a, J_b] = i \epsilon_{abc} J_c$ of the algebra su(2),  
where $a,b,c=1,2,3$, and
$\epsilon_{abc}$ is the totally antisymmetric tensor~\cite{ZFG}.
The generators $J_a$ of su(2) commute with the Casimir
operator $C \doteq J_1^2 +J_2^2 +J_3^2 \equiv J_4 (J_4 +1)$
which, in the Schwinger picture, leads to the identification
$J_4 \equiv (n_2 + n_1)/2 $. Consistently, one can check that
$[J_4, J_a] \equiv 0$.
Therefore, su(2) can be used to rewrite
Hamiltonian (\ref{HQ2}) which takes the nonlinear form
\bq
H = 2[ UJ_4^2 - v J_4 + UJ_3^2 - T J_1] \, .
\label{HJ2}
\eq
The spin picture embodies explicitly in $H$ the dimension $(2J+1)$
of the Hilbert space ${\cal H}(N)$ of physical states of $H$, where
the eigenvalue $J$($ =N /2$) of $J_4$ is the index of the spin
representation~\cite{ZFG}. The standard basis
${\cal B}_N =\{|J, m \rangle, |m|\le J=N/2 \}$
($J_3 |J, m \rangle = m |J, m \rangle$)
of ${\cal H}(N)$ is related to the number operator states through
$|J, m \rangle = |n_1, n_2 \rangle$, where $J= (n_2 + n_1)/2$,
$m = (n_2 - n_1)/2$. Thus $H$ and $J_a$
can be seen as $(2J+1) \times (2J+1)$ matrices.
The fact that $2 J_4 \equiv N$ is a constant of motion due to
$[H, {N}]=0$ is ensured by $[J_4, J_a] \equiv 0$.

An important consequence made explicit by the spin picture is
that the nonlinear term $J_3^2$ arising in Eq. (\ref{HJ2})
prevents the standard procedure of diagonalization of $H$
via a unitary trasformation of the group SU(2).
Such a procedure works only for matrices that can be written
as linear combinations of the algebra generators in that, by
construction, they can be reduced to one of the generators
by some appropriate group transformations~\cite{ZFG}.
Since the diagonalization is greatly simplified by such a
reduction, the latter practically identifies with the
diagonalization itself.
When a matrix $\cal O$ contains nonlinear terms of its generating
algebra (as in the case of Hamiltonian (\ref{HJ2}) with respect to
su(2)), resorting to a larger algebraic structure $\cal A$ enables one to
express $\cal O$ as a linear combination of generators of $\cal A$.
Such an enlarged algebra is called a
{\it dynamical algebra}~\cite{PER} for $\cal O$.

About Hamiltonian $H$ the problem is solved by
representing the algebra su$_{_N}$(2) [subscript $N (=2J)$
recalls the algebra link with the total
boson number] as a sub-algebra of $\cal A$ = su($M$)
for a suitable values of $M$.
In Appendix A we show that $\cal A$ $\equiv$ su$_{_1}$($N$+1).
The realization in $\cal A$ of the su$_{_N}$(2) generators
occurring in $H$ reads
$$
J_3 = \Sigma_q (J + 1 -q) n_{q}, \,
J_+ = \! \Sigma_q\, [q \, (2J+ 1 - q)]^{\frac{1}{2}}
\, b^+_q b_{q+1},
$$
$J_- \! \equiv (J_+)^+$ ($J_{\pm}= J_1 \pm iJ_2$),
where $b^+_q$ ($b_{q}$) are raising
(lowering) bosonic operators, $n_q = b^+_q b_{q}$,
and $q \in [1, N+1]$.
Recalling that the constraint $N_b:= \Sigma_q n_q \equiv 1$
must be accounted for (see Appendix A) we shall call the realization
just obtained {\it single-boson picture}.
This furnishes the simplest way to represent in a linear form the
nonlinear term of $H$. In fact, Hamiltonian (\ref{HJ2}) becomes
\bq
H  = C +
2U\! \Sigma_q [ m^2(q)\, n_{q} \!
- \tau R(q, J)\,(b^+_q b_{q+1}\! + h.c. )] \; ,
\label{HOB}
\eq
with $\tau := \! T/U$,
$R(q, J) :=\! [(J \!+\! 1/2)^2-(m(q)-1/2)^2]^{1/2}$,
$m(q) := J+1-q$, and $C=2 [UJ^2 \! -v J]$, where
the quadratic term has been re-expressed as a linear combination
of number operators $n_q$.
A valuable alternative form of the original model (\ref{HQ2})
is offered by Eq. (\ref{HOB}) which recasts the two-well dynamics
in terms of
the dynamics of a single boson on a linear nonhomogeneous lattice. 
In the single-boson picture physical
states are expressed as~\cite{COM4} 
\be
|\Psi \rangle = \Sigma_q \, \psi_q  \, b^+_q |0,...,0 \rangle
\label{STA}
\ee
with the normalisation condition $\Sigma_q |\psi_q|^2=1$
(see Appendix A).
The system dynamics thus takes place
on a hypersphere inside ${\bf C}^{N+1}$, where it can be interpreted
in a classical form. This matches the fact that states (\ref{STA})
can be shown to be su$_{_1}$($N$+1) coherent states~\cite{PER}.
Based on Eq. (\ref{STA}), the Schr\"odinger problem
$i \partial_t |\Psi \rangle = H |\Psi \rangle$
can be rewritten as a set of equations of motion
\bq
i {\dot \Psi}_m = 2 U m^2 \, \Psi_m -
T \left [R^{_{(J)}}_{m+1}\, \Psi_{m+1} + R^{_{(J)}}_{m}\, \Psi_{m-1}
\right ] \, ,
\label{PSI}
\eq
where we have introduced $\Psi_m := \psi_q$, and
$R^{_{(J)}}_m := R(q, J)$ with $m := J+1-q$, to join the
present formalism with the spin picture basis ${\cal B}_N$
where states are labeled by $m \in [-J, J]$. It is worth
noting how Eqs. (\ref{PSI}) can be derived as well from the
effective Hamiltonian
\bq
\langle H \rangle
\! = \!C + \!\! \Sigma_m [2U m^2 |\Psi_m|^2 \! -\! T \! 
R^{_{(J)}}_m
\left ( \Psi^*_m \Psi_{m-1} \!\!+ \!c.c. \right )] \; ,
\label{HJB}
\eq
representing~the~energy~expectation value
$\langle \Psi|H |\Psi \rangle \equiv \langle H \rangle$,
provided the Poisson
structure $\{ \Psi_m,\Psi^*_{\ell}\}=\delta_{m \ell}/i \hbar$ is
assumed (see commment~\cite{COM5}). The time evolution of the
dynamical variables (the components of $|\Psi \rangle$) is
determined once the initial condition
$|\Psi (0) \rangle$ at $t=0$ has been assigned.
Upon denoting by $X_m$ the components of the $H$ eigenstates $|X \rangle$,
one can retrieve the secular equation
\bq
E X_m = 2 U m^2 X_m -T \left [R^{_{(J)}}_{m+1}
\, X_{m+\!1} + R^{_{(J)}}_{m} \, X_{m-\!1}
\right ] ,
\label{SEC}
\eq
from Eq. (\ref{PSI}). $X_m $'s can be shown to be real.

The procedure relying on the dynamical algebra
construction has led to interpret model (\ref{HQ2})
as a lattice model with one boson via Eq. (\ref{HOB}).
This procedure has been implemented as well for illustrative purposes
since it shows clearly how the nonlinearity occurring
in the matrix form of $H$ is transferred to the coefficients
of the linear combination in $\cal A$. We point out that
such a simplification also works for a linear chain of $S$
interacting wells with $N$ bosons whose Hamiltonian can be written
via the generators of su$_{_N}$($S$). The latter, in fact,
can be always immersed within an algebra su($M$) with $M$
sufficiently larger than $S$; moreover the two-boson realization
of su($M$) can be obtained for any $M$. 

The component form of secular equation (\ref{SEC}) represents,
at the operational level, a basic intermediate result. The latter,
as shown in Sec. III, is used to
characterize explicitly (in the same spirit of Bloch's theorem for
electronic wave functions) the inner spectrum structure as well as
the structure of the energy eigenstates.
%
\section{spectrum structure}
In order to investigate the spectrum structure,
we consider first the effects of the PS and the OS
on the energy eigenstates, and make
explicit how such two symmetries strongly characterize the
eigenstate components. Then, we employ the results
of such an analysis to recast the eigenvalue equation related
to equation (\ref{SEC}) in a recursive form, and identify the
parameters able to control the splitting of the energy levels.
The case with half-integer $J$ and integer $J$ are treated
separately.

The PS is realized via the action of the unitary transformation
$U_1 := {\rm exp}[i \pi J_1]$ which takes $J_3 = (n_2 -n_1)/2$
into $U_1 J_3 U^+_1 =-J_3$. This matches the effect of the
PS classical action which involves the 
exchange of populations $n_1$ and $n_2$.
Let us introduce the hermitian operator $P :=$
${\rm exp}[-i\pi J] \, U_1$ whose action on the states
$|m \rangle$ (we drop the representation index
$J$ in $|J,m \rangle$ since it is fixed) is deducible from
the equation $U_1 |m \rangle = {\rm exp}[i\pi J] |-m \rangle$.
Owing to $[H , U_1 ]= 0$, $P$ can be diagonalized together with $H$.
The action of $P$ on the standard basis,
$P |m\rangle = |-m \rangle$, implies that, for a generic state
$|\Psi \rangle $,
$$
P: |\Psi \rangle = \Sigma_m \Psi_m |m\rangle
\to P |\Psi \rangle =
\Sigma_m \Psi_m |-m\rangle \,.
$$
In particular, the $P$ action on an eigenstate $|X \rangle$
entails the component transformation $X_m \to \sigma  X_{-m}$,
where $\sigma$ is not fixed, in Eqs. (\ref{SEC}).
Actually, these remain unchanged since
$R^{_{(J)}}_{m+1} \equiv R^{_{(J)}}_{-m}$
for each $m$. The fact that $P^2 \equiv {\bf I}$ fixes $\sigma$
showing how the allowed eigenvalues for $P$ are $\sigma= \pm 1$.  
Since $P: X_m \to \pm  X_{-m}$, each eigenstate has thus a definite
symmetry character under $m \to -m$.

This fact suggests to reorganize the vectors basis in terms of
vectors $|m,\pm \rangle =(|m\rangle \pm | -m\rangle)/ \sqrt{2}$
that are either symmetric or antisymmetric. The new basis allows
one to define in ${\cal H}(N)$ a symmetric (antisymmetric)
subspace ${\cal H}^+(N)$ (${\cal H}^-(N)$) whose vectors have
components we will denote by $\Phi^+$ ($\Phi^-$) such that
$P \Phi^{\pm}= \pm \Phi^{\pm}$. When the description in the new
basis is adopted then Eqs. (\ref{SEC}) for the eigenvector components
can be written in the matrix form
\be
E 
	  \left[
	    \begin{array}{c}
	      X^+ \\ 
	      X^-
	     \end{array}
	\right]  = 
   \left[
   	\begin{array}{ccc}
	S_J(T) & \vline & 0 \\
	\hline  
	0 &\vline & A_J(T)
	\end{array}
   \right] 
   	\left[
		\begin{array}{c}
		X^+ \\ 
		X^-
		\end{array}
	\right] \, ,
\label{SECN}
\ee
where $S_J(T)$ ($A_J(T)$) is the sub-matrix associated with 
symmetric (antisymmetric) sector, and $0$ represents the zero-matrix.
The matrix equation (\ref{SECN}) is separable in two independent
equations for $X^+$ and $X^-$: Their explicit form which depends on
the representation index $J$ is displayed in the sequel.

The odd symmetry (denoted by OS) is obtained by combining the
action $U_3 J_1\, U^+_3=-J_1$ of $U_3 := {\rm exp}[i\pi J_3]$ on
$-TJ_1$ in $H$ with the change $T \to -T$ which restores the
initial form of $H$. Since $J$ can be either integer or half-integer,
for considering the two cases separately
it is convinient to introduce the matrix
$L_{{J}} (T) :=
|| L_{m \ell}|| $ 
\bq
L_{m \ell}
= 2U  m^2 \delta_{m,\ell} 
- T [ R^{_{(J)}}_{m} \delta_{m,\ell+ 1} +
R^{_{(J)}}_{\ell} \delta_{m+1,\ell} ] ,
\label{COMMTX}
\eq
where
$m,\ell = {1/2}, {3/2},\ldots,J$ if $J$ is half-integer, and
$m,\ell = 1,2,\ldots,J$ if $J$ is integer.
%

Let us start with the half-integer case. In Eq. (\ref{SECN}),
the sub-matrices $S_{_J}(T)$ and $A_{_J}(T)$
coincide with the matrix $L_{_J}(T)$ up to the quantity
$-\eta T R^{_{(J)}}_{1/2}$ which must be added
to the matrix element
$L_{^{\frac{1}{2} \frac{1}{2}}}$ with $\eta=-1(+1)$
in the antisymmetric (symmetric) case. Representing $U_3$ in the
basis ($\{ |m,\pm \rangle \}$) entails
\be
U_3 = \left[
   	\begin{array}{ccc}
	0 & \vline & D \\
	\hline  
	- D &\vline & 0
	\end{array}
   \right] 
\label{U3}
\ee
in which $D=Diag(i,-i,\ldots)$. The action of $U_3$ on any vector
takes its symmetric components into the antisymmetric ones and vice
versa, namely $P U_3 \Phi^{\pm}= \mp U_3 \Phi^{\pm}$ if
$P \Phi^{\pm} = \pm \Phi^{\pm}$.
The structure of energy spectrum is reconstructed
through the following three observations:

\noindent
{\bf i}) The secular equation for $X^+$ and $X^-$ derived from
Eqs. (\ref{SECN}) can be written in the reduced form
\bq
E C_m \! =\! 2 U m^2 C_m - T \left [R^{_{(J)}}_{m+1} C_{m+\!1} +
R^{_{(J)}}_{m} C_{m-\!1} \right ] ,
\label{SER}
\eq
with $C= X^+, X^-$, for $1/2 < m \le J$, whereas
\bq
0 \! = [ 2 U(1/2)^2  -\!  \eta T R^{_{(J)}}_{1/2} - E] C_{1/2} \! 
-\! T R^{_{(J)}}_{3/2}\, C_{3/2} \, ,
\label{SPE}
\eq
holds for $m = 1/2$, where $\eta \equiv -1\, (+1)$ in the
antisymmetric (symmetric) case. For a given $T\neq 0$,
Eqs. (\ref{SER}) and (\ref{SPE}) show that the eigenvalues
$E$ can be seen as a set of functions
$E_a(T,{\eta})$ of $\eta$ with $1/2 \le a \le N/2$, defined
implicitly.
The energy eigenvalue $E_a(T, {+1})$
of each symmetric eigenstate can be derived from that $E_a(T, {-1})$
of an antisymmetric eigenstate by moving $\eta$ from $-1$ to $+1$,
and vice versa.

\noindent
{\bf ii}) We consider here the problem of ordering the set of
eigenvalues $E_a(T, {\pm 1})$.
For $T \to 0$ (this eliminates the $\eta$ dependence)
Eqs. (\ref{SEC}) are solvable; the resulting eigenvalues are doubly
degenerate since the equations for both the symmetric and the
antisymmetric components are identical.
Explicitly, $T \to 0$ entails $E_a(T,{+1}), E_a(T,{-1}) \to 2U m^2$
for some $m$ which shows how the label $a$ can be identified with
$m \in [1/2, J]$. The order induced by (positive) $m$ on the set
$\{ 2U m^2 : |m|\le J \}$ is inherited both by the symmetric
eigenvalues $\{E_a(T, +1) \}$ and by the antisymmetric
eigenvalues $\{E_a(T, -1) \}$ as proven by the limit $T \to 0$.
This also implies that, for sufficiently small $T$,
$E_a(T,{\pm 1}) \ne E_b(T,{\pm 1})$ if $a \ne b$.

\noindent
{\bf iii}) Implementing the action of $U_3$ whose matrix form is
given by Eq. (\ref{U3}) on Eq. (\ref{SECN}) leads to the equation  
\be
E 
	 \left[
	    \begin{array}{c}
	      \tilde{X}^- \\ 
	      \tilde{X}^+
	     \end{array} \!
	\right] \! = \!
   \left[
   	\begin{array}{ccc}
	A_J(-T) & \vline & 0 \\
	\hline  
	0 &\vline & S_J(-T)
	\end{array}
   \right] \!\!
   	\left[
		\begin{array}{c}
		\tilde{X}^- \\ 
		\tilde{X}^+
		\end{array}\!
	\right] \, ,
\label{SECNN}
\ee
where $\tilde{X}^{\mp} \! =\! D X^{\pm}$,
and $P\,\tilde{X}^{\pm}\! = \pm \tilde{X}^{\pm}$.
The substitution $T \to -T$ entails that
[we use the simplified notation $E^{\pm}_a(T):= E_a(T,\pm 1)$]
the set of the symmetric (antisymmetric)
eigenvalues $\{ E^+_a(T) \}$ ($\{ E^-_a(T) \}$) coincides with
the set of the antisymmetric (symmetric) ones $\{ E^-_b(-T) \}$
($\{ E^+_b(-T) \}$), where $1/2 \le a, b \le J$.
Notice that, in general,
$E^{\pm}_a(T) \equiv E^{\mp}_{b}(-T)$, where
not necessarily $b$ coincides with $a$. Nevertheless, for $T \to 0$
$E^{\pm}_a(0) \equiv E^{\mp}_{b}(0) \equiv 2U a^2= 2U b^2$
implies that $b=a$, as pointed out at point (${\bf ii}$).

As a consequence of points (${\bf i}$)-(${\bf iii}$),
we find that the symmetric eigenvalues are associated with the
antisymmetric ones through the formula
\be
E_a^\pm(T ) = E_a^\mp (-T) \; \; (1/2 \le a \le J) \, .
\label{PvT}
\ee
Also, since the eigenvalues equation can be cast in an iterative
form via the recurrence equation
\bq
{\sl d}_m(E) = \! (2Um^2-E) {\sl d}_{m\!+\!1}(E) \! -\! T^2
[R^{_{(J)}}_{m+\!1}]^2 {\sl d}_{m\!+\!2}(E) ,
\label{REC}
\eq
which starts from
\bq
0 =\left[{U}/{2} -\! E+ \eta T R^{_{(J)}}_{{1}/{2}}
\right]
{\sl d}_{\frac{3}{2}}(E) -
T^2 [R^{_{(J)}}_{{1}/{2}}]^2 {\sl d}_{\frac{5}{2}}(E) ,
\label{ESEP}
\eq
and terminates with ${\sl d}_J(E) = 2UJ^2- E$, consistently with
({\bf iii}) one finds that the eigenvalues cannot be even functions
of $T$.

For integer $J$, the dimension of matrix $S_{_J}(T)$ changes
from that of matrix $A_{_J}(T)$. In the antisymmetric case one
finds $A_J(T)=L_J(T)$, while in the symmetric
case ($S_J(T) := ||S_{m n}||$),
where the indices runs over $0,1,\ldots,J$, one finds
$S_{0 1} = S_{1 0}= -T R^{_{(J)}}_{\, 1}$,
$S_{m n} = L_{m n}$ for $m, n \ge 1$, and
$S_{m n} =0 $ otherwise.
Eqs. (\ref{SER}) still hold for integer $J$ provided $2 \le m \le J$.
The two special cases are those corresponding to $m= 0,1$  
\be
0= -E \, C_0 - T \sigma R^{_{(J)}}_{\, 1} \, C_{1}  \, ,
\label{SERI}
\ee
\be
0= (2 U  \, -E) \, C_1
-T \left [R^{_{(J)}}_{2} \, C_{2} + \sigma \,
R^{_{(J)}}_{0} \, C_{0}
\right ] \, ,
\label{SPEI}
\ee
with $C= X^+, X^-$.
The parameter $\sigma$ must be set equal to one in the symmetric
case ($C= X^+$), while in the antisymmetric case ($C= X^-$)
the expected elimination of the component $X^-_0$ follows from
setting $\sigma =0$. Hence the dimensions of Hilbert sub-spaces
are such that dim~${\cal H}^-(N)$= dim~${\cal H}^+(N)-1$,
while the secular equation for $X^-$
will have a degree diminished of one. Explicitly, one has
\bq
0 = \! \left[ \frac{E (2 U\! -\! E)}{T^2} \! +\!
\sigma^2 [R^{_{(J)}}_{\,1}]^2 \right] \!
{\sl d}_2(E) \!-\!  E [R^{_{(J)}}_{2}]^2  \! {\sl d}_3(E) .
\label{ESEN}
\eq
In the symmetric case a $(J+1)$-th degree equation for $E$
issues from (\ref{ESEN}) through formula (\ref{REC}). Comparing
the eigenvalue equations for the symmetric ($\sigma =1$) and
antisymmetric ($\sigma =0$) states shows that each, but one,
symmetric eigenvalue merges to an antisymmetric one when $\sigma$
goes from $0$ to $+1$.
Due to the diversity of the secular equation with $\sigma =1$
from that with $\sigma =0$, 
even in the case with integer $J$, the energy
spectrum is constituted
by $2J$ nondegenerate eigenvalues $\{ E_a^\pm (T):
1\le a \le J \}$ that for $T \to 0$ form $J$ pairs
$E_a^\pm (T) \to 2Ua^2$ and a single one $E_0 (T)$ which goes
to zero in the same limit.
Also, due to the quadratic dependence of
Eqs. (\ref{REC}), (\ref{ESEN}) on $T$,
Eq. (\ref{PvT}) must be replaced with
\be
E_a^\pm (T) = E_a^\pm (-T)  \quad (1 \le a \le J) \, ,
\ee
which, contrary to what happens with half-integer $J$,
maps each eigenvalue in itself under $T \to -T$.
In addition, of course, one must consider $E_0 (T)=E_0 (-T)$ as well.
Figs. 1 illustrate the spectrum structure dependence on $T/U$
for $N= 6,7$ (see also Fig. 2).

\section{discussion}

The interesting feature disclosed by the above analysis
is the possibility to recognize both in the half-integer and
in the integer case two inner parameters ($\eta$ and $\sigma$)
that control, in a way independent of $T$, the level splitting
generating the doublets.
The limit $T \to 0$ causes a coalescence of doublet levels such
that $E_a^{\pm} (T) \to 2Ua^2$ which
suggests $T$ as a possible perturbative parameter for evaluating
the level splitting.
On the other hand, Fig. 2 clearly shows that each
eigenvalue $E_a^{+} (T)$ remains close
to its partner $E_a^{-} (T)$ on a finite range $I_a(T)$
of $T$ indexed by the eigenvalue label $a$.
In the half-integer case, 
this implies that, inside $I_a(T)$, indeed $\eta$ represents
a good perturbative parameter
(recall that $E_a^{\pm} (T)= E_a(T,\eta)$ with $\eta =\pm 1$)
which allows one to evaluate $E_a(T,\pm 1)$ by
perturbing $E_a(T,\eta)$, e. g., around $\eta =0$.
For integer $J$, where the level separation is controlled
by $\sigma \in [0,1]$, one can observe an effect similar to that
showed in Fig. 2: the symmetric eigenvalue $E_a(T, 1)=E_a^+ (T)$
remains close to its antisymmetric partner $E_a(T, 0)= E_a^- (T)$
on a finite range. Because the function $E_a(T, \sigma)$ joins
analitically $E_a^- (T)$ to $E_a^+ (T)$ then $\sigma$ can be
assumed as the perturbative parameter for the present case.
 
The actual size of the range $I_a(T)$
can be evinced roughly from Fig. 2, where
the level separation strongly diverges only
when $E_a^\pm (T)$ cross $E \equiv E_* := NT$
(recall that $E_*$ is the energy critical value defined in the
introduction; its derivation can be found in Ref.~\cite{FPZ}).
This rule seems to be motivated from the
insensitivity of Eq. (\ref{ESEP}) (Eq. (\ref{ESEN}))
from the parameter $\eta$ ($\sigma$)
in the terms
$$
\eta T R^{_{(J)}}_{{1}/{2}}-E \quad
(\sigma^2 [R^{_{(J)}}_{\,1}]^2-E^2/T^2)\, ,
$$
for suitable values of $E$, $T$ and of the other coefficients.

Concerning the classical limit effected through $N \to \infty$,
numerical simulations with large boson numbers $N$ show that the
series of doublets becomes degenerate (coalescence of the doublet
levels) thus restoring the conditions that allow for
the SB effect. How recovering the latter is briefly
illustrated via the following comparison between the classical
and the quantum behavior of the two-well system. 
Classically, at a given energy $E > E_*$, the system
described in $\cal P$ evolves either
on $\gamma_-$ or on $\gamma_+$ (see Sec. I).
Trajectories $\gamma_-$ and $\gamma_+$ are
such that the populations' gap $n_2 - n_1$ weakly oscillates
around opposite values $-\mu$ and $+\mu$, respectively.
Quantally, for $E > E_*$, the combination
$|C_a^{\pm} \rangle = |X_a^{+} \rangle \pm |X_a^{-} \rangle$
of the symmetric/antisymmetric
eigenstates $|X_a^{\pm} \rangle$ of the $a$th doublet can be
shown to provide opposite nonvanishing expectation values
$\langle J_3 \rangle = \pm \chi$ of $J_3 = (n_2 - n_1)/2$.
This fact is caused by the eigenstate structure and was discussed 
in Ref.~\cite{FPZ}.
Then, states $|C_a^{\pm} \rangle$ can be associated naturally to
a pair of isoenergetic orbits $\gamma_-$, $\gamma_+$ that have
$\mu \equiv 2 \chi $. Increasing $N $, the time-dependent
mixed state 
\be
| \Psi\rangle = e^{-it E_a^+ (T)/\hbar} | X_a^+ \rangle +
e^{-it E_a^- (T)/\hbar} | X_a^- \rangle 
\label{MIX}
\ee
(satisfying the Schr\"odinger problem of $H$) exhibits
a sort of temporary self-trapping effect
(i. e. the localization either on $\gamma_-$ or on $\gamma_+$)
which is repeated periodically and has a duration increasing with $N$.
In fact, because of the oscillations of
the factor ${\rm exp}{ \{ [it (E_a^+ (T) -E_a^- (T)]\} }$ 
between $+1$ and $-1$ entailing
$| \Psi\rangle \propto | C_a^+ \rangle$ and
$| \Psi\rangle \propto | C_a^- \rangle$, respectively,
the system stays in a quantum state involving
the localization on $\gamma_{\pm}$
for intervals of the order of the period
$\Delta t = 2\pi \hbar/[E_a^+ (T) -E_a^- (T)]$
that increase when the level separation is reduced.
The system remains definitively in the classical-like states 
(full emergence of the SB effect inducing the
self-trapping)
when the tunneling time
from $| C_a^+ \rangle$ to $| C_a^- \rangle$ diverges, namely for
$[E_a^+ (T) -E_a^- (T)] \to 0$ (coalescence of doublet levels
induced by $N \to \infty$).

\section{conclusions}
In Sec. I we have reviewed the dynamics of coupled Bose condensates
described by Eqs. (\ref{CGP}) (in the approximation with zero
Laplacian terms) emphasizing the SB phenomenon that occurs in the
phase space when increasing the energy over the critical value $E_*$.
Such a phenomenon (and the ensuing self-trapping effect
on isoenergetic orbits $\gamma_{\pm} \in {\cal P}$
where $n_2 -n_1$ oscillates around nonzero values $\pm \mu$)
has prompted the study of
model (\ref{HQ2}) which represents the quantum counterpart
of model (\ref{CGP}) within the two-mode approximation of the
condensate field operator.
One of the purposes of the present work
was to investigate the dynamical behavior corresponding, at the
quantum level [through model (\ref{HQ2})], both to the SB effect
and to the related self-trapping. An aspect we have particularly
deepened is the quantum counterpart (level splitting) of the
bifurcation mechanism generating pair of isoenergetic trajectories
$\gamma_{\pm}$ when $E$ crosses $E_*$.

The formal set-up for studying the energy spectrum has been
developed in Sec. II by recasting model (\ref{HQ2}) for the
boson modes $a_1$, $a_2$ into the matrix form (\ref{HJ2})
within the spin formulation {\it \`a la} Schwinger.
The spin form of Hamiltonian (\ref{HJ2}) makes easily viable
the derivation of the secular equation (\ref{SEC}).
The latter has been achieved by using the dynamical algebra method
(this enacts systematically the reduction of Hamiltonian
nonlinearities) whose application is described in Appendix A.

In addition to supplying equation (\ref{SEC}),
the use of the dynamical algebra method has shown
the implicit link of the secular equation with
the effective-bosons model (\ref{HOB}).
Such a model reformulates the two-well dynamics of Hamiltonian
$H$ with $N$ bosons in a noticeably simplified form that consists
of a single boson hopping on a nonhomogenous lattice (single boson
picture).
The interest for the single boson picture and the underlying
formal construction is motivated by the possibility of extending
it to more structured models such as a chain model of
$S$ condensates with $N$ boson. The case $S=3$, which raises interest 
owing to its nonintegrable dynamical character, is presently under
investigation~\cite{FP2}.

Sec. III has been devoted to make explicit the structure of
the energy spectrum based on the symmetries of $H$.
Upon introducing the permutational symmetry (PS) and recognizing
the further odd symmetry (OS), we have employed them to characterize
the Hilbert space of $H$ as well as the energy eigenstates, both
for half-integer $J$ (odd number of boson $N$) and for integer $J$
(even number of boson $N$).
The $N$-dependent form of the secular equation obtained in the
two cases has led to the central result of Sec. III, namely
to recognize the possibility of introducing in a natural way
inner parameters that control the level
splitting in the energy spectrum doublets.

Such parameters --$\eta$ [$\sigma$] is defined in Eq. (\ref{ESEP})
[Eq. (\ref{ESEN})] for half-integer [integer] $J$--
have shown that the splitting originating the doublets can be
generated explicitly in an analytic way. This fact combined
with the doublets' structure exhibited in Fig. (2) suggests
that parameters $\eta$, $\sigma$ can be used as perturbative
parameters in approximating the doublet levels
inside the regions of the $T/U$ axis where the levels keep close.
This approach may be preferable than the standard
perturbative approach depending on the natural parameter $T/U$:
this, in fact, can be shown to require higher and higher powers
of $T/U$ when approximating levels far from the
ground-state~\cite{AFKO}.

We emphasize the fact that
generating the level splitting via the inner
parameters $\eta$, $\sigma$ can be interpreted as the quantum
form of the bifurcation mechanism issuing pairs of orbits
$\gamma_{\pm}$. In general, our construction should furnish the
quantum framework for describing the bifurcation effects of
any system whose Hamiltonian (in the critical regions of its
phase space) has locally the same form of $H$. In this sense,
both the quantum phase models and spin models~\cite{AMPE}
in the mesoscopic system physics promise interesting
applications.
We notice as well that the matrix/algebraic analysis underlying
the study of `quantum' bifurcation effects gives a valuable,
both formal and practical, tool for characterizing quantally the
chaos onset in the model with $S=3$.

The classical limit $N \to \infty$ has been
commented in Sec. IV, where the symmetry breaking (SB) effect
inherent in the classical two-well dynamics is recovered
from the quantum scenario via superposition
(\ref{MIX}) of the symmetric and antisymmetric states
of each doublet. In fact, the coalescence of the doublet levels
caused by $N \to \infty$ (revealed by numerical simulations)
leads, through a limiting process, to inhibit the oscillations
of the system state $|\Psi \rangle$ between $|C_a^{+} \rangle$ and
$|C_a^{-} \rangle$ which ends up by coinciding with one of such
states. This realizes the localization interpreted classically
as the self-trapping effect. 

We conclude by illustrating a possible reformulation
of Eq. (\ref{SEC}) in a continuous form valid for large $J$
directed to extimate the low part of energy spectrum.
Setting
$X_m \equiv$$Y_m / [(m+\! J)! (J-\! m)!]^{1/2}
$
in Eq. (\ref{SEC}) provides
$(2 U m^2 \! -\! E)\! Y_m =
T [(J-\! m)Y_{m+\! 1}\! + \! (J+\! m) Y_{m-\! 1}]$,
which reduces to the second order equation (${\dot F}:= dF/dx$)
\be
\ddot{F} -2xs {\dot F}
+ (2J+1-s R+E/T) F/R =0 \, ,
\label{CONT}
\ee
where $s:=\pm 1$ and $x := m/\sqrt{JR}$ with $R^2 :=1+2UJ/T$, while
$Y(m) \equiv {\sl exp} [-\alpha m^2/J]\,
{F}(m/ \sqrt{JR}\,)$, with $\alpha := (sR-1)/2$,
is the dependence of the (rescaled) components $Y_m$ on
(the continuous variable) $m$. 
The assumption that ${F}$ identifies with the
$n$-th Hermite polynomial gives the equation
$2n \equiv[ 2J+1-R+E/T]/R$ for $s=+1$ which, in turn,
supplies a set of eigenvalues. Their dependence on $T$ is
compared in Fig. 3 with the lowest part of the spectrum of
$N =20$ bosons. This result as well as the results/observations
discussed above have prompted further work that is
presently in progress.

\acknowledgments
We are indebted to E. Guadagnini and M. Rasetti for stimulating
discussions. The financial support of I.N.F.M. (Italy) and of
M.U.R.S.T. (within the Project Sintesi) is gratefully acknowledged.

\begin{appendix}
\section{}

The type of enlarged algebra we deal with is $\cal A$ = su($M$)
[the latter can be viewed as the generalized version
of the spin algebra with $M^2-1$ generators].
Selecting an appropriate value of $M$ allows one to rewrite the
nonlinear Hamiltonian $H$ in terms of a linear combination of
generators of $\cal A$.
This furnishes $\cal A$ with the status of dynamical algebra for $H$.
To construct explicitly $\cal A$ it is useful
to consider the two-boson form of the su(M) generators
$
E_{ij}:= b_i^+ b_j , \, (i \ne j)
$,
$
H_{ij}:= (b_i^+ b_i - b_j^+ b_j)/2
$
that satisfy the commutators~\cite{ZFG}
$$
[E_{ij}, E_{kl}]=\delta_{jk} E_{il}
- \delta_{il} E_{kj} \, ,
$$
$$
[E_{ij}, H_{kl}]= ( \delta_{jk} E_{ik}
-\delta_{jl} E_{il} +\delta_{il} E_{lj}- \delta_{ik} E_{kj} )/2 \, ,
$$
with $1 \le i, j, l, k \le M$.
Within the present realization of su($M$),
the representation theory of semi-simple Lie
groups~\cite{PER} states that
the eigenvalue $Q$ of the invariant operator
$N_b =\Sigma_i b_i^+ b_i$ ($[N_b, E_{ij}]=0$) selects 
a specific representation of su($M$) [$N_b$ can be viewed
as the total particle number relatively to the creation
(destruction) processes caused by $b_i^+$ ($ b_i$)].
In fact, the dimension of the Hilbert space basis ${\sl B}(M, Q) =
\{ |n_1, ...,n_M \rangle, Q= \Sigma^{_M}_{^{i=1}} n_i \}$
is given by
$$
dim\, {\sl B}(M,Q) = (Q +M-1)!/[(M-1)! Q\,!]\, ,
$$
where the states of the basis ${\sl B}(M,Q)$ are defined as
$|n_1, ..., n_M \rangle= \otimes^{_M}_{i=1} |n_i \rangle $ and the
number operator states $|n_j \rangle$ fulfil the equations
$$
b_i|n_i \rangle =\! {\sqrt{n_i}} |n_i-\! 1 \rangle, \;
b^+_i |n_i \rangle = \! {\sqrt{n_i +\! 1}} |n_i+\! 1 \rangle \, .
$$
The simplest realization of $\cal A =$ su$_{_Q}$($M$) is achieved by
setting $Q=1$ (single-boson picture); combining this fact with
the requirement of preserving the dimension $N+1$ of ${\cal H}(N)$,
entails $dim\, {\sl B}(M,1) = M  \equiv N+1$ thus providing
$\cal A =$ su$_{_1}$($N$+1) as a dynamical algebra for $H$.
The states of the related basis
$\{ |q \rangle = b^+_q |0,...,0 \rangle, q \in [1, N+1] \}$
are in a one-to-one correspondence
with the states $|J; m \rangle$, $|m| \le J= N/2$,
of the su$_{_N}$(2) standard basis
($J_3 |J; m \rangle = m |J; m \rangle$) via the index
map $m =\! J+\! 1-\! q$.

\end{appendix}
\vskip 0.3 truecm
%



\end{multicols}
\newpage
\noindent

\begin{figure}[htbp]
\label{fig1}
\caption{First figure shows the energy eigenvalues vs. $T/U$ for
$J=7/2$; second figure shows the case $J=3$ (different types of line
identify different eigenvalues).}
\end{figure}

\begin{figure}[htbp]
\label{fig2}
\caption{Solid lines represent the eigenvalues $E^{\pm}_a$
vs. $T/U$ for $J=15/2$. The dashed line describes the curve $E=2J T$. 
For each pair $E^{\pm}_a$, the splitting becomes evident when
$2 JT $ reachs $E^{\pm}_a$.}
\end{figure}

\begin{figure}[htbp]
\label{fig3}
\caption{The figure shows the seven lowest eigenvalues vs. $T/U$
for $J=10$ (solid lines); the seven dashed lines corrispond to
the approximate eigenvalues derived from Eq. (\ref{CONT}) for $s=+1$.
The approximation remains good for large values of $T/U$.}
\end{figure}


\begin{thebibliography}{000}
\bibitem{COM2}
Geometrically, one can say that
$\cal P$ exhibits a bifurcation of its periodic orbits and a
separatrix manifold that splits it into three regions:
$C_0$ with orbits invariant under the PS, and $C_{\pm}$ whose
(noninvariant) orbits are in ono-to-one correspondence via the
PS action.
\bibitem{COM4}
It is important to recall that the system dynamics is
embodied in the components $\psi_q$. The boson operators  
$b^+_q$, $b_q$ as well as the basis $\{ b^+_q |0,...,0 \rangle \}$  
have a purely formal role involved by the enlargement of the
algebra and of the Hilbert space thereof.
\end{thebibliography}
\end{document}